# Discrete Boltzmann Modeling of Compressible Flows

Aiguo Xu[1,2]  Guangcai Zhang[1]  Yudong Zhang[1,3,4]

1  Institute of Applied Physics and Computational Mathematics, Beijing 100088, China
2  Center for Applied Physics and Technology, MOE Key Center for High Energy Density Physics Simulations, College of Engineering, Peking University, Beijing 100871, China
3  Key Laboratory of Transient Physics, Nanjing University of Science and Technology, Nanjing 210094, China
4  Graduate School, China Academy of Engineering Physics, Beijing 100088, China

**Abstract:** Mathematically, the typical difference of Discrete Boltzmann Model (DBM) from the traditional hydrodynamic one is that the Navier-Stokes (NS) equations are replaced by a discrete Boltzmann equation. But physically, this replacement has a significant gain: a DBM is roughly equivalent to a hydrodynamic model supplemented by a coarse-grained model of the Thermodynamic Non-Equilibrium (TNE) effects, where the hydrodynamic model can be and can also beyond the NS. Via the DBM, it is convenient to perform simulations on systems with flexible Knudsen number. The observations on TNE are being obtaining more applications with time.

## 1. Introduction

Generally, compressible flow is frequently referred also to as gas dynamics which is the branch of fluid mechanics that deals with flows having significant changes in fluid density. That is because gases, mostly, display such behaviour. In fact, all materials are compressible. Besides liquids, gases and plasmas, to some extent, the plastic solids under strong shock can be modelled as compressible flows, too. Because, in the last case, that the strength of material is negligibly smaller than that of the shock. Flows with a Mach number less than 0.3 are usually treated as being incompressible for that the variation of density due to velocity is less than 5% in that case. The study on compressible flow is relevant to various fields, such as high-speed aircraft, rocket motors, jet engines, high-speed entry into a planetary atmosphere, gas pipelines, etc. where shock wave and/or detonation play a significant role. In this chapter the following several kinds of flows, including high Mach number flows with combustion, multiphase flows with phase separation, and complex flows with hydrodynamic instability,[1] are taken as examples, and all flows are treated uniformly as being compressible. It is hopeful that the methods and ideas developed in this chapter may be adapted, more or less, to other kinds of complex flows.

Some common and typical features of these flows are as below: Each of them possesses multi-scale structures and/or kinetic modes. There exist plenty of interfaces inside the system. The interfaces include material interfaces and mechanical interfaces. Each of them experiences complicated competitions between behaviours in various spatial-temporal scales. The forcing and responsive processes inside the system are very complicated. Such a system generally shows pronounced non-equilibrium behaviours.

## 1.1. Traditional models

The traditional macroscopic models of compressible flows are generally based on Navier-Stokes (NS) equations, even Euler equations. The model of Euler equations assumes that the system is always at its Local Thermodynamic Equilibrium (LTE). The NS model considers the Thermodynamic Non-Equilibrium (TNE) via the viscous stress and heat flux. The viscous tress and heat flux compose a set of convenient and effective description of the TNE. But the description is also quite dense and coarse-grained. Many specific TNE behaviours are invisible under NS description, even though they are helpful for understanding the specific TNE status. Besides, since includes only the first order term of Knudsen number in the Chapman-Enskog expansion[1], it is reasonable only when the Knudsen number is very small. It cannot be used to access deeper TNE behaviours. To access the complicated non-equilibrium behaviours, one possible solution is to use the Molecular Dynamics (MD)[2] or Direct Simulation Monte Carlo (DSMC)[3]. The MD simulation can help to understand some fundamental mechanisms from the atomic level. But the spatial and temporal scales it can access are too small to be comparable with experiments. The DSMC simulation has a similar constraint of computational cost.

The NS model is not enough to capture so complex non-equilibrium behaviours while the MD and DSMC cannot access spatial-temporal scales large enough. Under such conditions, a kinetic approach based on the Non-equilibrium Statistical Mechanics (NESM)[1, 4, 5] is preferable.

---

[1] It is clear that there exist significant overlaps among those classifications.

## 1.2. Non-Equilibrium Statistical Mechanics

Statistical mechanics is a branch of theoretical physics. It was developed to study the average behaviour of a mechanical system with uncertain state by using probability theory. Microscopic mechanical laws do not contain concepts such as temperature, heat, or entropy; however, statistical mechanics shows how these concepts arise from the natural uncertainty about the state of a system when that system is prepared in practice. Statistical mechanics provides exact methods to connect thermodynamic quantities to microscopic behaviour.

The NESM is based on mechanics and some necessary assumptions. The concept of macroscopic observation and assumption of coarse-grained density are cornerstones. The Liouville equation[4, 5] is the most fundamental governing equation when without quantum fluctuations. It describes the $N$-particle system using a partial-differential equation for the probability density function, $f = f(\xi_1, \xi_2, \cdots, \xi_N, t)$, in a $6N$-dimensional space,

$$\frac{\partial f}{\partial t} + \sum_{i=1}^{N} \left( \frac{p_i}{m} \cdot \frac{\partial}{\partial q_i} - \frac{\partial}{\partial q_i} \Phi^N \cdot \frac{\partial}{\partial p_i} \right) f = 0, \tag{1}$$

where $m$ is the particle mass, $\zeta_i = (q_i, p_i)$, $q_i$ and $p_i$ are the coordinate and momentum of the $i$-th particle, $\Phi^N(q_1, ..., q_N)$ is the interaction potential of all the $N$ particles. By integration over part of the variables, the Liouville equation can be transformed into a chain of equations, which is referred to as the Bogoliubov-Born-Green-Kirkwood-Yvon (BBGKY) hierarchy[4], where the first equation connects the evolution of one-particle probability density function with the two-particle probability density function, the second equation connects the two-particle probability density function with the three-particle probability density function, and generally the $S$-th equation connects the $S$-particle probability density function, $f_s$, with the (S+1)-particle probability density function, $f_{s+1}$. Specifically,

$$\frac{\partial f_S}{\partial t} + \sum_{i=1}^{S} \left( \frac{p_i}{m} \cdot \frac{\partial}{\partial q_i} - \frac{\partial}{\partial q_i} \Phi(q_1, ..., q_s) \cdot \frac{\partial}{\partial p_i} \right) f_S = \frac{N-S}{V} \int \sum_{i=1}^{S} \left( \nabla_{r_i} \Phi_{i,S+1} \right) \cdot \left( \nabla_{p_i} f_{S+1} \right) d^6 \zeta_{S+1}, \tag{2}$$

where $V$ is the volume of system, $\Phi(q_1, ..., q_s) = \sum_{i<j \leq s} \Phi_{i,j}(q_i, q_j)$ is the total two-particle interaction potential, $\Phi_{i,S+1}$ is the interaction potential between the particle $i$ and particle $S+1$. Here

$$V^n = \int_\Gamma f_n(\zeta_1, \zeta_2, \cdots, \zeta_n, t) d^6\zeta_1 \cdots d^6\zeta_n, \tag{3}$$

and

$$f_S(\zeta_1, \zeta_2, \cdots, \zeta_S, t) = V^{(S-N)} \int f_N(\zeta_1, \zeta_2, \cdots, \zeta_N, t) d^6\zeta_{S+1} \cdots d^6\zeta_N. \qquad (4)$$

When $S = N$, the above equation recovers to the Liouville equation. For a macroscopic system, the particle number, $N$, is in the order of Avogadro constant, $10^{23}$. Generally, there is no way to solve Eq. (1) or Eq. (2).

We need simplify the model via considering some simpler cases. If considering only the case where correlations among three and more particles are negligible, the two particle interaction is relevant to their distance, and the two particle distribution function can be written as the product of two single particle distribution functions, specifically,

$$\Phi_{1,2}(\boldsymbol{q}_1, \boldsymbol{q}_2) = \Phi_{1,2}(|\boldsymbol{q}_1 - \boldsymbol{q}_2|), \qquad (5)$$

$$f_2(\boldsymbol{q}_1, \boldsymbol{p}_1; \boldsymbol{q}_2, \boldsymbol{p}_2; t) = f_1(\boldsymbol{q}_1, \boldsymbol{p}_1; t) f_1(\boldsymbol{q}_2, \boldsymbol{p}_2; t), \qquad (6)$$

we obtain the Boltzmann Equation[1, 4],

$$\frac{\partial f}{\partial t} + \boldsymbol{v} \cdot \frac{\partial f}{\partial \boldsymbol{r}} + \boldsymbol{a} \cdot \frac{\partial f}{\partial \boldsymbol{v}} = Q(f, f), \qquad (7)$$

with

$$Q(f, f) = \int_{-\infty}^{+\infty} \int_{0}^{4\pi} \left( f^* f_1^* - f f_1 \right) g \, \sigma \, d\Omega \, d\boldsymbol{v}_1. \qquad (8)$$

Here is the single particle distribution function. Compared with the MD and Liouville equation, the Boltzmann equation is a much coarser-grained model. The purposes of establishing Boltzmann equation are to define and calculate the entropy in non-equilibrium state, to derive, prove, even modify the fundamental hydrodynamic differential equations[5].

## 1.3. Non-Equilibrium Statistical Mechanics and Macroscopic description

All hydrodynamic quantities are some kinds of kinetic moments of the distribution function and can be expressed as:

$$\boldsymbol{W} = \int f \begin{bmatrix} 1 \\ \boldsymbol{v} \\ \boldsymbol{v} \cdot \boldsymbol{v}/2 \end{bmatrix} d\boldsymbol{v}, \qquad (9)$$

where the particle mass has been assumed to be 1, $\boldsymbol{W}$ is the vector of macroscopic quantity composed of density, momentum, and energy, $\boldsymbol{W} = [\rho, \rho\boldsymbol{u}, \rho E]^T$. Besides, the total stress $\boldsymbol{\sigma}$, viscous stress $\boldsymbol{\Pi}$ and heat flux $\boldsymbol{q}$ are related to $f$ via

$$\sigma = \int f\ (v-u)(v-u)dv = \Pi + pI, \tag{10}$$

$$\Pi = \int (f - f^{eq})\ (v-u)(v-u)dv, \tag{11}$$

and

$$q = \frac{1}{2}\int (f - f^{eq})(v-u)(v-u)\cdot(v-u)dv, \tag{12}$$

respectively. Compared with Boltzmann equation, in Navier-Stokes equations,

$$\frac{\partial \rho}{\partial t} + \nabla \cdot (\rho u) = 0, \tag{13}$$

$$\frac{\partial (\rho u)}{\partial t} + \nabla \cdot (pI + \rho uu) = -\nabla \cdot \Pi, \tag{14}$$

$$\frac{\partial}{\partial t}(\rho E) + \nabla \cdot [(\rho E + p)u)] = -\nabla \cdot [q + \Pi \cdot u], \tag{15}$$

where $\Pi = 2/3\mu(\nabla \cdot u)I - \mu(\nabla u)^T - \mu\nabla u$, $q = -\kappa\nabla T$, $T$ is the temperature, $\mu$ and $\kappa$ are the coefficients of viscosity and heat conductivity, respectively. More details of molecular motions are neglected and the distribution function $f$ disappears as well. What remained are the conserved kinetic moments, density $\rho$, momentum (density) $\rho u$ and energy(density) $\rho E$, and a few non-conserved quantities, momentum fluxes $\rho uu$ and $\Pi$, energy fluxes $\rho Eu$ and $q$. The relation between internal energy $\rho T$ and pressure $p$ is given by the equation of state. In Euler equations, the viscous stress $\Pi$ and heat flux $q$ are further neglected.

From molecular dynamics to Boltzmann equation, to Navier-Stokes, and further to Euler equations, in each step, the description becomes coarser-grained, and the contained physical information becomes less[6].The switching of model in each step corresponds to the state of system under consideration gets closer to its thermodynamic equilibrium, the behaviour is simpler, consequently the system can be described by fewer physical variables. For a given non-equilibrium system, the switching of model in each step corresponds to the spatial-temporal scale that we use to observe the system becomes larger, consequently more smaller structures and quicker kinetic modes are invisible. What obtained are the remaining larger structures and slower kinetic modes. Based on Boltzmann equation, the most relevant TNE effects accompanying the hydrodynamic behaviours can be studied, in addition to the general hydrodynamic behaviours described by the hydrodynamic model.

## 2. Discrete Boltzmann theory

### 2.1. Discrete Boltzmann Modeling

From Boltzmann equation to DBM, two steps of coarse-grained physical modelings are needed. The principle for coarse-grained modeling is that the physical quantities used to measure the system must keep the same values after simplification.

**Step 1.** Linearization of the collision term

Even though, compared with MD or Liouville equation, Boltzmann equation is a much coarser-grained model, its collision term is still too complicated to be solved in most practical cases. The simplest way to simplify is to introduce a local equilibrium velocity distribution function, $f^{eq}$, and write the collision term into the following linear form,

$$\frac{\partial f}{\partial t} + \boldsymbol{v} \cdot \frac{\partial f}{\partial \boldsymbol{r}} + \boldsymbol{a} \cdot \frac{\partial f}{\partial \boldsymbol{v}} = -\frac{1}{\tau}(f - f^{eq}). \tag{16}$$

The physical meaning of the linearized collision model is thus: collisions of molecules result in that $f$ approaches to thermodynamic equilibrium $f^{eq}$ and the relaxation time is controlled by the parameter $\tau$. If we do not aim to measure the system using the specific values of $f$, but using only some of its kinetic moments, then only if these kinetic moments keep invariable after the simplification, that will be OK. The linearization of the collision term requires

$$\int -\frac{1}{\tau}(f - f^{eq})\boldsymbol{\Psi} d\boldsymbol{v} = \int Q(f,f)\boldsymbol{\Psi} d\boldsymbol{v}, \tag{17}$$

where $\boldsymbol{\Psi} = [1, \boldsymbol{v}, \boldsymbol{vv}, \boldsymbol{vvv}, \cdots]^T$ contributes the kinetic moments used to measure the system. The specific form of $f^{eq}$ depends on the terms that $\boldsymbol{\Psi}$ takes. According to the specific form of $\boldsymbol{\Psi}$ or $f^{eq}$, the linearized collision model may be referred to as the Bhatnagar-Gross-Krook (BGK) model[7-10], ellipsoidal statistical BGK model[11], Shakhov model (for monatomic gas)[12], Rykov model (for diatomic gas)[13], Liu model[14], etc. When only the mass, momentum and energy conservation laws are kept, $f^{eq}$ takes the simplest form, the Maxwellian. This is the so-called BGK model. Due to its simplicity, the BGK model is most extensively used.

It should be note that the Single-Relaxation-Time model works when all the kinetic modes approaching to thermodynamic equilibrium share more or less the same relaxation time. For more complicated cases where the relaxation times of different kinetic modes approaching to thermodynamic equilibrium are significantly different, the Multiple-Relaxation-Time (MRT) collision model is needed[15].

**Step 2.** Discretization of the particle velocity space

We first consider the case without the force term. The discrete Boltzmann equation reads,

$$\frac{\partial f_i}{\partial t} + v_{i\alpha} \cdot \frac{\partial f_i}{\partial r_\alpha} = -\frac{1}{\tau}\left(f_i - f_i^{eq}\right), \tag{18}$$

where $i$ is the index of discrete velocity. Since the common schemes for discretizing the space and time do not work for discretizing the particle velocity space. To find an effective means to discretize the particle velocity space, we go back to consider what we really need. Here, we do not aim to describe the system using specific values of the discrete distribution function $f_i$, but its kinetic moments. So, only if these kinetic moments, originally in integral form, can be equally calculated in summation form, that will be acceptable. Specifically,

$$\int f \, \boldsymbol{\Psi}'(\boldsymbol{v}) d\boldsymbol{v} = \sum_i f_i \, \boldsymbol{\Psi}'(\boldsymbol{v}_i), \tag{19}$$

where the left hand side gives the kinetic moments of $f$ needed to describe the system and $\boldsymbol{v}_i$ at the right hand side is the discrete particle velocity. According to the Chapman-Enskog analysis, a kinetic moment of $f$ can be finally calculated via an appropriate kinetic moment of $f^{eq}$. Therefore the requirement of Eq.(18) can be further written as

$$\int f^{eq} \, \boldsymbol{\Psi}''(\boldsymbol{v}) d\boldsymbol{v} = \sum_i f_i^{eq} \boldsymbol{\Psi}''(\boldsymbol{v}_i), \tag{20}$$

where the left hand side gives the kinetic moments of $f^{eq}$ being involved in the process of constructing DBM. The requirement of Eq. (19) can be rewritten as a matrix equation,

$$\hat{\boldsymbol{f}}^{eq} = \boldsymbol{M} \cdot \boldsymbol{f}^{eq}, \tag{21}$$

where $\hat{\boldsymbol{f}}^{eq} \equiv [\hat{f}_k^{eq}]$, $\boldsymbol{f}^{eq} \equiv [f_i^{eq}]$, $\boldsymbol{M} \equiv [m_{ki}]$, $\hat{f}_k^{eq}$ is the $k$-th kinetic moment of $\boldsymbol{f}^{eq}$. The elements, $m_{ki}$, are determined by the discrete velocities. That is to say, the discrete velocities should be chosen in such a way that the requirement of Eq. (20) is satisfied. Under such a constraint, the discretization of particle velocity space is flexible. In fact, $\boldsymbol{M}$ can also be non-squared rectangular matrix. The choice of discretization scheme depends on the compromise between the numerical cost, stability, and physical gain (such as physical symmetry).

Via the same idea, it is straight forward to formulate MRT-DBM. To ensure the relaxation times have clear physical meanings, in the MRT-DBM, the collision term is first calculated in the kinetic moment space, and then transformed back to the discrete velocity space. To ensure DBM describe reasonable flow behaviours, a correction term is needed[15].

According to the Chapman-Enskog analysis, to access systems which is deeper into thermodynamic non-equilibrium, higher order terms in Knudsen number should be considered. As a result, the requirement of $f$ in Eq. (18) will lead to more kinetic moment relations of $f^{eq}$ in Eq. (19). Consequently, more discrete velocities are needed. For the case with inter-particle force, one should generally first finish the derivative calculation of $f$ with respect to $\mathbf{v}$, and then perform the discretization of particle velocity space.

### 2.2. Non-equilibrium: definition and measuring

Once the concept of equilibrium is defined, the concept of non-equilibrium is clear. In classical mechanics, a particle is in mechanical equilibrium if the net force on that particle is zero. By extension, a physical system made up of many parts is in mechanical equilibrium if the net force on each of its individual parts is zero. In addition to defining mechanical equilibrium in terms of force, there are many alternative definitions for mechanical equilibrium which are all mathematically equivalent. In terms of momentum, a system is in equilibrium if the momentum of its parts is all constant. In terms of velocity, the system is in equilibrium if velocity is constant. In a rotational mechanical equilibrium the angular momentum of the object is conserved and the net torque is zero. More generally in conservative systems, equilibrium is established at a point in configuration space where the gradient of the potential energy with respect to the generalized coordinates is zero. Similarly, a fluid system is in fluid mechanical equilibrium or hydrodynamic equilibrium if the net force on each of its 'fluid particles' (small fluid elements) is zero and without temperature gradient around the 'fluid particle'.

For ideal gas system, in thermodynamic equilibrium there are no net macroscopic flows of matter or energy either within a system or between systems. In non-equilibrium systems, by contrast, there are net flows of matter or energy. Global thermodynamic equilibrium means that the relevant intensive parameters are homogeneous throughout the whole system, while local thermodynamic equilibrium means that those intensive parameters are varying in space and time, but are varying so slowly that, for any point, one can assume thermodynamic equilibrium in some neighbourhood about that point. Rarefied gases at ordinary temperatures behave very nearly like ideal gas and the Maxwell speed distribution is an excellent approximation for such gases. Thus, it forms the basis of the kinetic theory of gases.

It is evident that Euler equations are used to investigate fluid flows which are at local thermodynamic equilibrium but mechanical non-equilibrium, while NS, Burnett and Super-Burnett equations are used to investigate fluid flows which are at mechanical and thermodynamic non-equilibrium. Only when all kinds of kinetic moments of $(f - f^{eq})$ are zero, the system is at thermodynamic equilibrium. Hydrodynamic Non-Equilibrium(HNE) drives the evolution of $f$ and results in viscous stress and heat flux.

In many practical cases, it is neither necessary to know all the details of $f$ nor necessary to know all the kinetic moments of $(f - f^{eq})$. Therefore, one can (i) care only the main feature of $f$, specifically, neglect the high order terms of Knudsen number, (ii) care only the kinetic moments of $(f - f^{eq})$ which are most relevant to the macroscopic behaviours under consideration, specifically, those involved in constructing the discrete Boltzmann model.

A centrally important motivation of DBM is to check, measure and analyze the non-equilibrium state and effects[6, 16-18]. The DBM presents two sets of measures for the TNE. One set is dynamically from the difference of the kinetic moments of $f$ and $f^{eq}$,

$$\boldsymbol{\Delta}_m = \boldsymbol{M}_m(f) - \boldsymbol{M}_m(f^{eq}), \tag{22}$$

where $\boldsymbol{M}_m(f)$ is the $m$-th rank moment of velocity tensor of $f$, $\boldsymbol{\Delta}_m^*$ can be obtained when $\boldsymbol{M}_m(f)$ is the $m$-th rank central moment. The other set includes the viscous stress and heat flux. The former describes the specific TNE status, while the latter describes the influence of those TNE to the macroscopic control equations. The former is local, while the latter is non-local. The former is finer, while the latter is coarser. At each step of DBM simulation, both the $f$ and $f^{eq}$ are calculated. Therefore, the TNE effects are naturally included in each step. The captured TNE effects are just those being most relevant to the macroscopic flow behaviours under consideration.

Entropy production is a highly concerned quantity in both physics and engineering studies. From the physics side, it is helpful for understanding the complex non-equilibrium behaviours. From the engineering side, a process with lower entropy production may have higher energy transformation efficiency. Following the way of defining entropy equilibrium equation in the non-equilibrium thermodynamics, a new entropy equilibrium equation can be obtained as follows[19],

$$\frac{\partial s}{\partial t} = -\nabla \boldsymbol{J}_s + \sigma, \tag{23}$$

where $s$ is the entropy density,

$$\boldsymbol{J}_s = s\boldsymbol{u} + \frac{1}{T}\boldsymbol{\Delta}_{3,1}^* \tag{24}$$

and

$$\sigma = \boldsymbol{\Delta}_{3,1}^* \nabla \frac{1}{T} - \frac{1}{T}\boldsymbol{\Delta}_2^* : \nabla \boldsymbol{u} + \rho \frac{Q}{T} F(\lambda) \tag{25}$$

are the entropy flux and entropy production rate, respectively. $F(\lambda) = \dot{\lambda}$ is a rate function describing the process of chemical reaction occurring in the system, where $\lambda$ is defined as the ratio of the product density to the overall density. From Eq. (24) one can find that the entropy production results from three kinds of resources, the Non-Organized Energy Flux (NOEF), Non-Organized Momentum Flux (NOMF), and chemical reaction. From the relation, one can study the various mechanisms resulting in increase of entropy and their relative importance[19].

The TNE behaviours are very complex and difficult to quantitatively investigate. Finding a convenient and efficient method to characterize the TNE status and effects is the corner stone. DBM presents such a potential approach[16-25].

## 2.3. DBM versus CFD

The traditional CFD needs to first know the exact form of the hydrodynamic equations, then design numerical algorithm according to the properties of those equations. The DBM is a coarser-grained model derived from the Boltzmann equation. In principle, it can be formulated and applied to simulate flows without knowing the exact form of the hydrodynamic equations, only if necessary kinetic moments of $f^{eq}$ are followed. From the perspective of physical application, a DBM is roughly equivalent to a hydrodynamic model supplemented by a coarse-grained model of TNE. Via the DBM, it is easy to perform multi-scale simulations in a wide range of Knudsen number.

## 3. Applications

## 3.1. Combustion system

Combustion has long been playing a dominant role in the transportation and power generation. To improve combustion efficiency and decrease pollution, in recent years, some new combustion concepts have been proposed. For example, pulsed detonation engine, spinning detonation engine, microscale combustion, nanopropellent, partially premixed and stratified combustion, plasma assistant combustion, cool flames, etc. All these new combustion concepts involve complicated non-equilibrium chemical and transport processes[26].

The chemical reaction process is very complex and may include varieties of reaction mechanism. So far, most of the chemical reaction kinetic models are phenomenological. As the first step, we consider only the simplified form of Lee-Tarver chemical reaction rate law[21]. Considering the thermal initiation, the reaction kinetic is described by

$$\frac{d\lambda}{dt} = \begin{cases} a(1-\lambda) + b(1-\lambda)\lambda, & T \geq T_{th} \text{ and } 0 \leq \lambda \leq 1 \\ 0, & \text{else,} \end{cases} \qquad (26)$$

where $a$ and $b$ are constants, $\lambda$ is the concentration of the product and works as the reaction process parameter, $T_{th}$ is the temperature threshold for chemical reaction. Consider the case where the chemical reaction is slow enough compared with the kinetic process of approaching thermodynamic equilibrium, so we can treat $f$ as $f^{*eq}$ during the reaction process. The evolution equation of single-relaxation-time DBM for combustion reads

$$\frac{\partial f_i}{\partial t} + v_{i\alpha}\frac{\partial f_i}{\partial r_\alpha} = -\frac{1}{\tau}(f_i - f_i^{*eq}), \qquad (27)$$

where $f_i^{*eq} = f_i^{eq}\left(\rho, u, T + (\gamma-1)\tau Q F(\lambda)\right)$ is the local equilibrium distribution function taking into account chemical reaction effect. $Q$ is the amount of heat released by the chemical reactant per unit mass. If the relaxation times of different kinetic modes approaching to thermodynamic equilibrium are significantly different, one needs the MRT model[22]. Some observations brought by DBM are as below.

Non-equilibrium quantities defined in Eq. (21) are used to study a simple case of detonation[21]. For the case of CJ detonation shown in Figure 1(a), the corresponding non-equilibrium quantities, $\Delta_2^*$ and $\Delta_3^*$, are shown in Figure 1(b) and 1(c), respectively. Interestingly, at the position of von Neumann pressure peak, the system is not far from but is nearly in its thermodynamic equilibrium. The internal energies in two degrees of freedom deviate from their mean value in opposite direction with the same amplitude. The internal energy in each degree of freedom deviates from the mean value in opposite direction before and after the von Neumann pressure peak. The amplitude in front of the von Neumann pressure peak is larger. The physical reasons are as below. The mechanical non-equilibrium is the driving force for TNE. When a strong shock comes, the density, temperature and flow velocity increases abruptly so that the system does not have enough time to relax to its thermodynamic equilibrium. With decrease of the gradients of density, temperature and flow velocity, the system gets relatively more time for thermodynamic relaxation. At the position of von Neumann pressure peak, the system has been close to its thermodynamic equilibrium. After the von Neumann pressure peak, the density, flow velocity decreases quickly so that the system does not have enough time for thermodynamic relaxation again. The total deriving force for TNE in front of the von Neumann pressure peak is larger than behind.

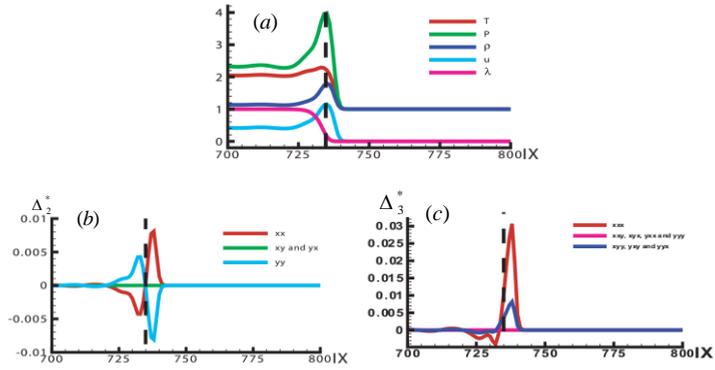

**Figure 1.** Profiles of physical quantities for CJ detonation. (a) is for the density $\rho$, pressure $P$, temperature $T$, x-component of velocity $u$, and the reaction process $\lambda$. (b) and (c) are for the non-equilibrium quantities $\Delta_2^*$ and $\Delta_3^*$, respectively ( See Ref. [21] for more details ).

Figure 2 gives the pressure P and corresponding non-equilibrium quantity $\Delta_{2,xx}$ at a certain time[22]. From Figure 2(a) to 2(c), the relaxation time, $1/R_i$, decreases by 10 times in each case so that the detonation wave changes from unsteady to steady. In Figure 2(d)-2(f), the shaded area, enclosed by the curve $\Delta_{2,xx}$ and the x-axis, presents a rough description on the global TNE effect around the detonation wave. It is clear that the viscosity (and/or heat conductivity) decreases the local TNE while increases the global TNE around the detonation wave.

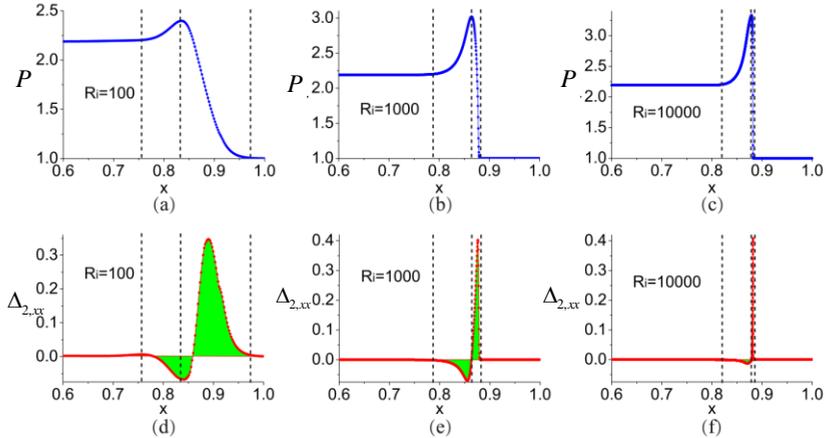

**Figure 2.** Profiles of physical quantities for different relaxation times. (a)-(c) are for the pressure $P$. (d)-(f) are for the corresponding non-equilibrium quantities $\Delta_{2,xx}$ ( From Ref. [22] with permission ).

Figure 3 shows some numerical results aiming to investigate the main mechanisms resulting in entropy increase and their relative importance in the combustion system[19]. It is clear that, in the checked cases, the most pronounced contribution to entropy increase is from the chemical reaction, $\Delta s_{CHEM}$, the lest contribution is from the non-organized energy flux, $\Delta s_{NOEF}$, the contribution of non-organized momentum flux, $\Delta s_{NOMF}$, is in between. With the increasing of Mach number, the entropy production caused by non-organized momentum flux becomes more remarkable.

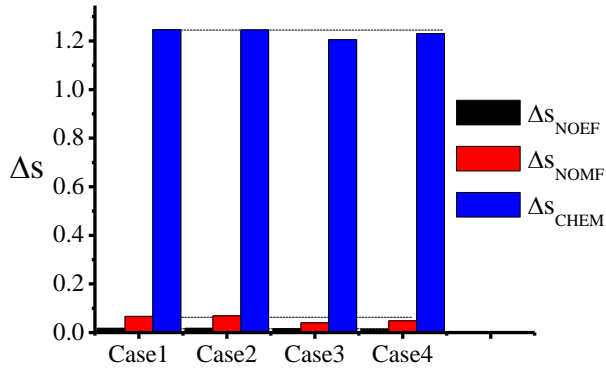

**Figure 3.** Mechanisms for global entropy production in four cases (From Ref. [19] with permission).

### 3.2. Multiphase flow with phase separation

Phase separation is an important branch in the field of multiphase flows. It is also a kind of non-equilibrium phase transition. The key step for modeling phase separation is to incorporate the non-ideal gas effects into the discrete Boltzmann equation. Enskog equation can be regarded as an extension of Boltzmann equation under the hard-ball molecule model[1]. Although the specific treatments may be different, the aims are the same. Those are all to replace the equation of state of ideal gas with a more practical one.

In 2007 Gonnella, Lamura, and Sofonea (GLS)[27] introduced an appropriate interparticle interaction to the Watari's model[28] to describe van der Waals fluids. The evolution equation of GLS model reads:

$$\frac{\partial f_{ki}}{\partial t} + v_{ki\alpha} \frac{\partial f_{ki}}{\partial r_\alpha} = -\frac{1}{\tau}(f_{ki} - f_{ki}^{eq}) + I_{ki}, \qquad (28)$$

where the external force term $I_{ki}$ takes the following form:

$$I_{ki} = -[A + B_\alpha(v_{ki\alpha} - u_\alpha) + (C + C_q)(v_{ki\alpha} - u_\alpha)^2] f_{ki}^{eq} \quad (29)$$

In a recent work[20], the GLS model was further developed to be a kinetic model which can be used to access both the hydrodynamic non-equilibrium and the thermodynamic non-equilibrium. To roughly and averagely estimate the derivation amplitude from the thermodynamic equilibrium, a TNE strength can be defined as

$$D = \sqrt{\Delta_2^{*2} + \Delta_3^{*2} + \Delta_{3,1}^{*2} + \Delta_{4,2}^{*2}} \quad (30)$$

Figure 4 shows that the maximum value point can work as a physical criterion to discriminate the two stages, spinodal decomposition and domain growth, of phase separation. The TNE strength increases with time in the first stage while decreases with time in the second stage. More details are referred to the original publication[20].

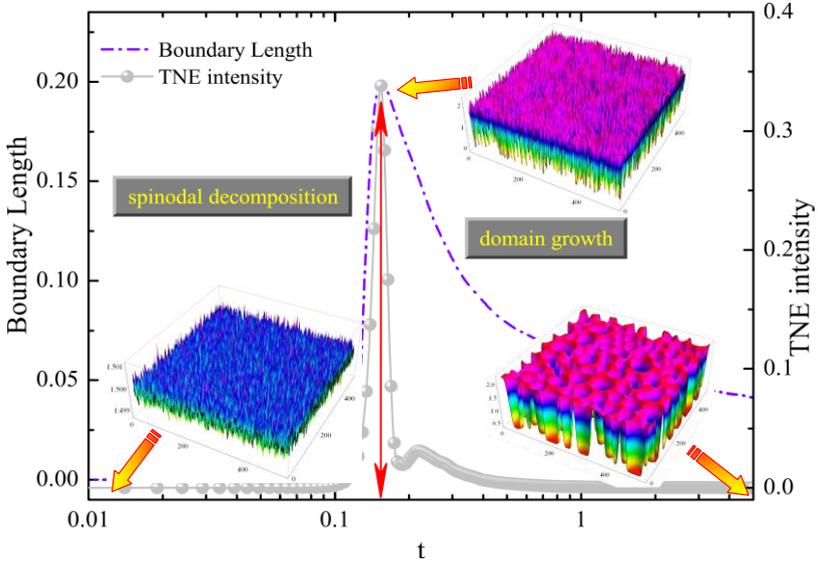

**Figure 4.** Evolutions of the boundary length $L$ and the TNE strength $D$ for the phase separation process (See Ref. [20] for more details).

### 3.3. Rayleigh-Taylor interfacial instability

Rayleigh-Taylor instability (RTI) occurs at the interface between two fluids with different densities. The compressible RTI system can be described by[23, 24]

$$\frac{\partial f_i}{\partial t} + v_{i\alpha} \frac{\partial f_i}{\partial r_\alpha} - \frac{a_\alpha(v_{i\alpha} - u_\alpha)}{RT} f_i^{eq} = -\frac{1}{\tau}(f_i - f_i^{eq}) \quad (31)$$

where $a_\alpha$ is the acceleration due to external force.

Generally, the depth of the mixing layer is an important parameter to measure the evolution of RTI. For incompressible RTI, the measurement is readily performed by tracing the constant density. However, for the compressible case, how to measure the mixing layer remains a thorny problem. Here we present two independent interface-tracking methods. One is by tracking the mean temperature of the upper and bottom fluids while the other is by tracking the maximum values of TNE characteristic quantities, such as $\Delta^*_{3.1,y}$. The latter method is based on the fact that $\Delta^*_{3.1,y}$ takes its maximum value at the position of the interface along the $y$ direction of the spike and bubble. The perturbation amplitudes developing with time obtained by the two methods are shown in Figure 5. The good agreement shows that the two approaches validate each other and the local TNE, $\Delta^*_{3.1,y}$, can be used to track interfaces in numerical experiments.

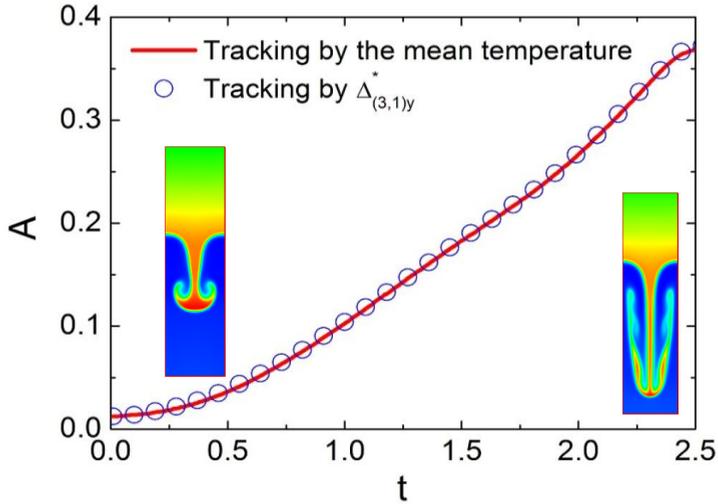

**Figure 5.** The perturbation amplitudes developing with time obtained by two different tracking approaches (See Ref. [23] for more details).

### 3.4. Compressible flows under shock

Figure 6 shows the profiles of physical quantities in the process where the shock wave passes outwards from the heavy medium to the light one. Figures 6 (a) and 6(b) are for the case without and with initial perturbations at the interface, respectively. From left to right, one can find three kind of interfaces, the rarefaction wave, material interface and shock wave, which are indicated by dashed lines[25].

According to the TNE information, the main feature of actual particle velocity distribution function can be qualitatively recovered. Figure 7 shows an example, where the interface is not perturbed initially. The details are referred to Ref.[25]. DBM simulations[25] show that the shear stress exists only for the oblique shocking. As a consistent correspondence, MD results[29] show that fluctuating shear stresses exist if observe in a scale with a few angstroms, while their mean value becomes negligibly small when being averaged over a scale with several hundred angstroms.

A comparison of the DBM results and those of the MD is shown in Figure 8. Figures 8(a) and 8(b) shows the DBM results for the cases where the interface is not and is perturbed initially, respectively. Figure 8 (c) shows the shear stresses from MD simulation. Since the MD uses particle description, the existence of locally fluctuating shear stress corresponds to the observation in the case with perturbed interface; the observation that mean value of shear stresses becomes negligibly small in a much larger scale roughly correspond to the case with non-perturbed interface for the DBM simulations.

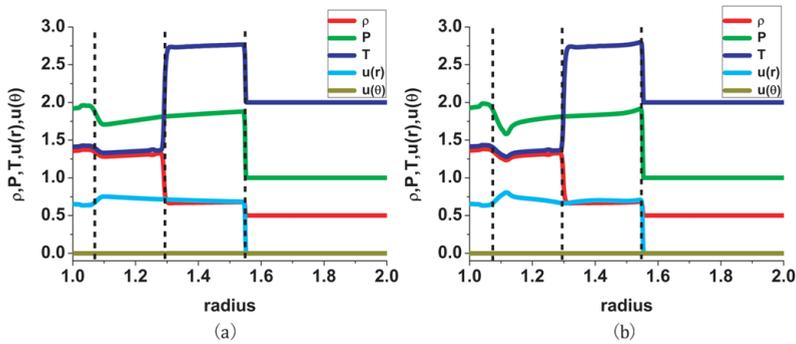

**Figure 6.** Profiles of physical quantities in the process of shock wave passing outwards from the heavy medium to the light one. (a) Without initial perturbations at the interface. (b) With initial sinusoidal perturbation at the interface (From Ref. [25] with permission).

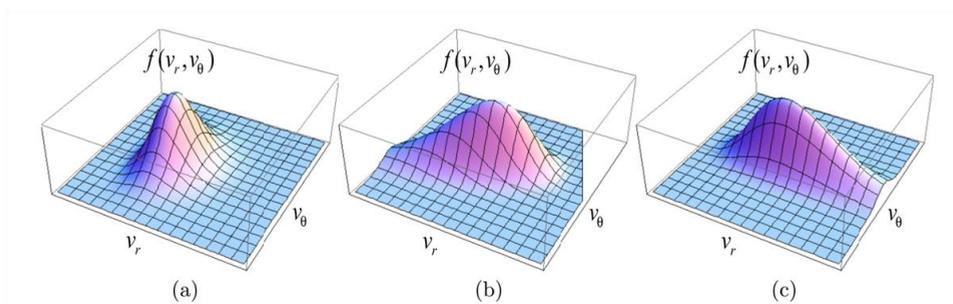

**Figure 7.** Sketches of the actual distribution functions in velocity space $(v_r, v_\theta)$. Panels (a)-(c) show the recovered distribution functions at the rarefaction wave, the material interface, and the shock wave, respectively (From Ref. [25] with permission).

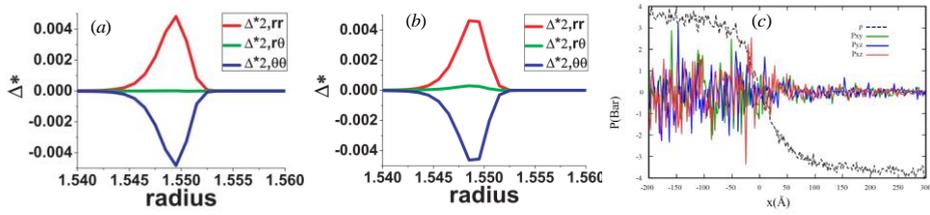

**Figure 8.** Shear stress within shock wave. (a) DBM results for the case without initial interface perturbation. (b) DBM results for the case with initial interface perturbation. (c) MD results (Figures (a)-(b) are from Ref. [25] with permission, Figure (c) is from Ref. [29] with permission).

### 3.5. Shock wave in plasma

Figure 9 shows an example for that the TNE effects can used to physically discriminate shock wave in plasma from those in common fluid. From the first two rows, the two TNE quantities, $\Delta_2^*$ and $\Delta_{4,2}^*$, show quite similar behaviours around shock wave and/or detonation wave in common fluid, even though they two have different physical meanings. However, the two quantities show qualitative difference around shock wave in plasma[30].

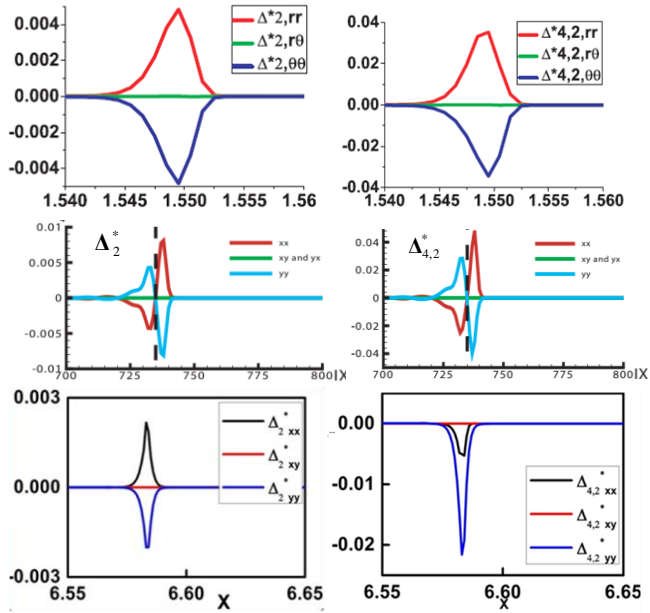

**Figure 9.** TNE characteristics of three types of shock waves. The three rows, from top to bottom, correspond to pure shock wave (See Ref. [25] for more details), shock wave with chemical reaction

(detonation) (See Ref. [21] for more details), and shock wave in plasma (See Ref. [30] for more details). The profiles on the left column show the values of $\Delta_2^*$ around the wave fronts while the profiles on the right column show the values of $\Delta_{4,2}^*$.

## 4. Summary

Understanding compressible flows needs more time. DBM presents a convenient way to model and simulate systems with trans-scale Knudsen number. Mathematically, the only difference of discrete Boltzmann from the traditional hydrodynamic modeling is that the NS equations are replaced by a discrete Boltzmann equation. But physically, this replacement has a significant gain: a DBM is roughly equivalent to a hydrodynamic model supplemented by a coarse-grained model of the TNE, where the hydrodynamic model can be and can also beyond the NS. The TNE provided by DBM has been used to investigate non-equilibrium effect during detonation process, to discriminate different stages of phase separation, to recover actual particle velocity distribution function qualitatively, to track the interfaces of different fluid, and to discriminate shock wave in plasma from those in common fluid. More use of those TNE quantities are further being discovered with the deeper investigate of the compressible and complex flows.

**Acknowledgement:** We warmly thank Profs. Wei Kang, Zhihua Chen, Yanbiao Gan, Feng Chen, Chuandong Lin, Huilin Lai, Zhipeng Liu, Bo Yan, et al. for helpful discussions. We acknowledge support of the National Natural Science Foundation of China(under Grant Nos. 11475028 and 11772064), Science Challenge Project (under Grant No. JCKY2016212A501 and TZ2016002), and Science Foundation of Laboratory of Computational Physics.